\documentstyle[12pt]{article}
\newcommand{\be}{\begin{equation}}
\newcommand{\ee}{\end{equation}}
\begin{document}
\begin{titlepage}
\setcounter{page}{1} 
\title{Does the non--minimal coupling of the scalar field improve or
destroy inflation~?} 
\author{Valerio Faraoni\\ \\
{\small \it Inter--University Centre for Astronomy and Astrophysics}\\
{\small \it Post Bag 4, Ganeshkhind, Pune 411~007, India}
} \date{} 
\maketitle
\thispagestyle{empty} 
\vskip1truecm 
\begin{abstract} 
The non--minimal coupling of a scalar field to the Ricci curvature in a
curved spacetime is unavoidable according to several authors.
The coupling constant $\xi$ is
not a free parameter: the prescriptions for the coupling
constant $\xi$ in specific scalar field and gravity theories (in
particular in general relativity) are studied. The results  are applied to the
most popular inflationary scenarios of cosmology and their theoretical
consistence is analysed.  Certain observational constraints on $\xi$
are also discussed.
\end{abstract} 
\vskip1truecm \begin{center} 
To appear in {\em Proceedings of the 7th Canadian Conference on General
Relativity and Relativistic Astrophysics} (Calgary, Canada 1997).
\end{center} 
\end{titlepage} 
\clearpage

\section{Non--minimal coupling of the scalar field}

The generalization of the flat space Klein--Gordon equation to a curved
space, 
\begin{equation} \label{1} \Box \phi-\xi
R \phi -\frac{dV}{d\phi}=0 \; , 
\end{equation} 
includes the possibility of
an explicit coupling between the scalar field $\phi$ and the Ricci
curvature of spacetime $ R $. A non--minimal coupling has been advocated
and is unavoidable in a quantum theory of $\phi$: it is generated by
quantum corrections even if it is absent in the classical action, or it is
required in order to renormalize the theory. This leads us to ask whether
physics selects a unique value (or a range of values) for the coupling
constant $\xi$.  What is the value of the coupling constant $\xi$~? This
question is relevant for different areas of theoretical physics, and it is
crucial for the application to cosmological inflation (see Ref.~\cite{PRD}
and references therein for an overview), the success of which is deeply
affected by the value of $\xi$. 
 The answer to the question depends on the underlying theory of gravity and 
of the scalar field. A relativist's answer is:\\ \\
{\em if gravity is described by a metric theory and the scalar
field $\phi$ has a non--gravitational origin}\footnote{An example of a 
scalar field with
gravitational origin is the Brans--Dicke field, or its generalization in 
scalar--tensor theories.}{\em , and satisfies Eq.~(\ref{1}), then} $ \xi=1/6
$ (``conformal coupling''). \\ \\
This result, proved in Ref.~\cite{SonegoFaraoni} and later confirmed 
in Refs.~\cite{GribPoberii}, was derived during the study of wave
propagation 
and tails of radiation in curved spaces, and it arises by imposing the
Einstein Equivalence Principle (EEP \cite{Will}) on the 
physics of the field $\phi$ (the structure of tails of radiation 
becomes closer and closer to that occurring
in flat spacetime when the curved manifold is progressively approximated by its
tangent space).
The result is completely independent of conformal transformations, the
conformal structure of spacetime, the spacetime metric $g_{ab}$ and the field
equations for $g_{ab}$; it is, however, unclear why the value of $\xi$ that
emerges from this analysis is precisely the one that gives conformal 
coupling. 
A naive explanation that can be given is the following: no preferred
length or mass scale is present in the flat space Klein--Gordon equation,
and therefore no such scale must appear in the corresponding curved space
equation for the massless field when small regions of spacetime are
considered, if the EEP holds.

The EEP holds in all metric theories of gravity,
and it must be imposed on $\phi$ if $\phi$ is a non--gravitational
field\footnote{If $\phi$ has gravitational origin, statements about its physics
pertain to the Strong Equivalence Principle, which is believed to be satisfied
only in general relativity \cite{Will}.}.
If $\xi \neq 1/6$ there is, in principle, the possibility that a {\em massive}
scalar field propagate along the light cones in a space in which $R\neq 0$.

A particle physicist's answer to the problem of the value of $\xi$ is 
different and more varied: 
\begin{itemize}

\item if $\phi$ is a Goldstone boson in a theory with spontaneously broken 
global symmetry, $\xi=0 $ \cite{VoloshinDolgov}; 

\item in the large $N$ approximation to the Nambu--Jona--Lasinio model, 
$ \xi=1/6$ \cite{HillSalopek};

\item in Einstein's gravity with backreaction and $V=\lambda \phi^3$, $  \xi=0$ 
\cite{Hosotani};

\item if $\phi$ is the Higgs field of the standard model $  \xi \leq 0$ or 
$\xi \geq 1/6$ \cite{Hosotani}; 

\item in any theory formulated in the Einstein conformal frame $\xi=0$ (this
includes supergravity, the low energy limit of superstring theories, 
Kaluza--Klein, and virtually all theories involving a dimensional
reduction and compactification of the extra dimensions
\cite{MagnanoSokolowski}).

\end{itemize}

Moreover, $\xi$ is subject to renormalization, apparently 
leaving little room for an
unambiguous determination of its value. Fortunately, in cosmology the 
prospects for the determination of the value of $\xi$ are better than it 
appears in the general case.

\section{Applications to cosmological inflation}
                 
The simplification occurs because inflation is a low energy, classical
phenomenon: the tensor contributions to the quadrupole in the cosmic microwave
background imply that the energy density of the inflaton 60 e--folds before 
the end of inflation satisfies  
$$V_{60} \leq 6\cdot 10^{-11} m_{pl}^4   \; , $$
where $ m_{pl} $ is the Plank mass. Hence, gravity was classical 
during inflation. The
inflaton field is decomposed into its unperturbed value plus quantum 
fluctuations (that seed density perturbations giving rise to galaxies, clusters
and superclusters later in the history of the universe): 
\begin{equation}  \label{2}
\phi=\phi_0 (t)+ \delta \phi (t, \underline{x}) \; .
\end{equation} 
The distribution of $\phi_0 $ is peaked around
classical trajectories and the evolution of $\phi_0$ is described by classical
equations (see e.g. Refs.~\cite{classinfl}). 

Cosmologists seem to have two different approaches to the problem 
of the value of $\xi$: most authors ignore the problem altogether by 
setting $\xi=0$ arbitrarily. Other authors use $\xi$ as a free parameter
that can be tuned {\em a posteriori} 
to minimize the troubles of specific inflationary scenarios. 
Instead, the value of the coupling constant is fixed {\em a priori} 
in many cases. We analysed the proposed inflationary scenarios by 
answering the questions:
\begin{itemize} 
\item is any prescription for $\xi$ applicable~?
\item what are the consequences of this prescription for the 
viability of the scenario~?  
\end{itemize}
The aspects taken into account are the existence of inflationary 
solutions, a sufficient amount of inflation to solve the problems of the
standard big bang model, the fine--tuning of initial conditions, and the
evolution of density perturbations. The results of Ref.~\cite{PRD} 
are summarized in
the following (partial) lists, in which ``theoretical consistence''
refers only to the value of $\xi$ employed in the specific gravity and
inflaton
theory used -- independent arguments may rule out the scenario. \\ \\
{\bf Theoretically consistent scenarios}
\begin{itemize} 

\item Power--law inflation (``PLI'') in any theory formulated in the 
Einstein frame

\item Extended inflation (original formulation and recast as PLI) 

\item Hyperextended inflation (original formulation and recast as PLI) 

\item Induced gravity inflation

\item Natural inflation

\item Double field inflation (original formulation and  version 
of Ref.~\cite{Copelandetal94}) 

\end{itemize}
\noindent {\bf Theoretically inconsistent scenarios}

\begin{itemize}

\item New inflation

\item Chaotic inflation (general relativity and $V=\lambda \phi^4 $) 

\item Chaotic inflation (general relativity and $V=\mu^2 
\left[ \phi^2/2 +\lambda/(2n) \phi^{2n} \right]$) 

\item Chaotic inflation (general relativity and 
$V=\lambda \left( \phi^2-v^2 \right)^2$) 

\item Double field inflation (version of Ref.~\cite{GarciaLinde95}). 
\end{itemize}

\section{Observational constraints on $\xi$}

Adopting a different point of view, it is desirable to constrain  the value
of the coupling constant $\xi$ by using the available observations of the
cosmic microwave background. Unfortunately, despite the fact that some
inflationary solutions are known for $\xi \neq 0$, very few predictions have
been made for the observables quantitites, in particular for the spectral index
$n_s$ of density perturbations. For chaotic inflation with the potential
$V=\lambda \phi^4$, one has \cite{Kaiser} 
\begin{equation}  \label{3}
n_s=1-\frac{32\xi}{1+960 \xi} \; .
\end{equation} 
The combined statistical analysis of the {\em COBE} and Tenerife
observations yields the 1$\sigma$ limits $ 0.9 \leq n_s \leq 1.6$, which 
imply 
\begin{equation}  \label{4}
\xi \leq -1.56 \cdot 10^{-3} \; , \;\;\;\;\;\;\;\;\;\;  \xi \geq -9.87 \cdot
10^{-4}        \; .
\end{equation} 
The value predicted by general relativity is $n_s=0.967$. For 
chaotic inflation with
the potential $V=\lambda \left( \phi^2-v^2 \right)^2 $, $n_s$ depends 
on $\xi $ only to second order in the relevant parameters:
\begin{equation}  \label{5}
n_s=1-\frac{2}{60+\pi \left( v/m_{pl} \right)^2} \; .
\end{equation}
The scalar field potential 
\begin{equation} \label{6} 
V( \phi)=\lambda \phi^n \;\;\;\;\; , \;\;\;\;\; n>6 
\end{equation}               
with $\xi \neq 0$ gives power--law inflation in the following regions of the 
$(n,\xi)$ parameter space:
\begin{equation}\label{7}
n>6 \;\;\;\; , \;\;\;\; 0<\xi <\frac{2}{n^2-12n+44} \; ,
\end{equation}
\begin{equation}  \label{8}
6<n<4+2\sqrt{3}\simeq 7.464 \;\;\;\; , \;\;\;\;\xi<0 \; ,
\end{equation}
\begin{equation}  \label{9}
n=4+2\sqrt{3} \;\;\;\; , \;\;\;\;\xi < \frac{1}{4(3-\sqrt{3})}\simeq 0.197
  \; ,                                      \end{equation}
\begin{equation}  \label{10}
n> 4+2\sqrt{3}  \;\;\;\; , \;\;\;\;\frac{-2}{n^2-8n+4}<\xi<0 \; .
\end{equation}
The range of values $6\leq n \leq 10$ is interesting 
for superstring theories; only a very narrow range of values of
$\xi$ is allowed for high $n$. 

\section{Conclusions}

In most inflationary scenarios, the coupling constant $\xi$ is not a free
parameter that can be tuned arbitrarily, but its value is fixed by the theory
of gravity and of the scalar field adopted. Moreover, the theoretical 
consistence of
many inflationary scenarios is deeply affected by the value of $\xi$. Some
scenarios turn out to be theoretically inconsistent, while others are viable 
according to the correct use of non--minimal coupling.
The feeling from our and from previous works (\cite{PRD} and references
therein) is that, in general, non--minimal coupling makes it harder to 
achieve inflation.

Work in progress includes studying the consequences of the
prescriptions of $\xi$ for the cosmic no--hair theorems,
calculations of density perturbations spectra for $\xi \neq 0$, and the
consequence of the general relativity--as--an--attractor behaviour during
inflation. Non--cosmological applications of the theory described here
include the physics of boson stars and of classical/quantum wormholes.

\vskip1.5truecm
\section*{Acknowledgments}

The author would like to acknowledge M. Bruni and S. Sonego  for
helpful discussions.

   \end{document}